\documentclass[reprint,twocolumn]{revtex4-1}

\usepackage[utf8]{inputenc}
\usepackage{amsfonts,amsmath,amssymb}
\usepackage{graphicx}
\usepackage{xspace}
\usepackage{mathtools}
\usepackage{hyperref}
\usepackage{lmodern}
\usepackage{microtype}

\let\oldcite\cite
\renewcommand{\cite}{\unskip~\oldcite}
\newcommand{\reftable}[1]{Table~\ref{#1}}
\newcommand{\reffig}[1]{Fig.~\ref{#1}}

\newcommand{\refsec}[1]{Sec.~\ref{#1}}
\newcommand{\refcite}[1]{Ref.\cite{#1}}

\newcommand{\see}[1]{(see \eg\refcite{#1})}

\newcommand{\gev}{\ensuremath{\,\text{GeV}}}
\newcommand{\tev}{\ensuremath{\,\text{TeV}}}
\newcommand{\flux}{\ensuremath{/\text{GeV}/\text{s}/\text{sr}/\text{m}^2}}
\newcommand{\intflux}{\ensuremath{/\text{s}/\text{sr}/\text{m}^2}}

\newcommand{\eg}{e.g.,\xspace}
\newcommand{\ie}{i.e.,\xspace}

\DeclarePairedDelimiter\parentheses{\lparen}{\rparen}

\newcommand{\probsymbol}{\ensuremath{p}}
\newcommand{\prob}[1]{\ensuremath{\probsymbol\parentheses*{#1}}}
\newcommand{\cond}[2]{\ensuremath{\prob{#1 \bm\mid #2}}}

\newcommand{\pvalue}{\ensuremath{\text{\textit{p}-value}}\xspace}

\makeatletter
\g@addto@macro\bfseries{\boldmath}
\makeatother

\begin{document}

\title{DAMPE squib? Significance of the $1.4\tev$ DAMPE excess}

\author{Andrew Fowlie}
\affiliation{ARC Centre of Excellence for Particle Physics at the Tera-scale, School of Physics and Astronomy, Monash University, Melbourne, Victoria 3800 Australia}

\date{\today}

\begin{abstract}
We present a Bayesian and frequentist analysis of the DAMPE charged cosmic ray spectrum. 
The spectrum, by eye, contained a spectral break at about $1\tev$ and a monochromatic excess at about $1.4\tev$. 
The break was supported by a Bayes factor of about $10^{10}$ and we argue that the statistical significance was resounding. 
We investigated whether we should attribute the excess to dark matter annihilation into electrons in a nearby subhalo. 
We found a local significance of about $3.6\sigma$ and a global significance of about $2.3\sigma$, including a two-dimensional look-elsewhere effect by simulating 1000 pseudo-experiments.
The Bayes factor was sensitive to our choices of priors, but favoured the excess by about $2$ for our choices.
Thus, whilst intriguing, the evidence for a signal is not currently compelling.
\end{abstract}

\maketitle

\section{Introduction}

The Dark Matter Particle Explorer (DAMPE) experiment recently published the energy spectrum of electrons and positions from about $10\gev$ to about $4\tev$\cite{Ambrosi:2017wek}. The spectrum, by eye, contained two interesting features: a break at about $1\tev$ and a monochromatic excess at about $1.4\tev$. The DAMPE analysis itself contained no statistical analysis of the excess, which, nevertheless, stirred much interest\cite{Liu:2017obm,Ding:2017jdr,Yang:2017cjm,Cao:2017sju,Ghorbani:2017cey,Nomura:2017ohi,Gu:2017lir,Zhu:2017tvk,Li:2017tmd,Chen:2017tva,Chao:2017emq,Niu:2017hqe,Gao:2017pym,Jin:2017qcv,Cholis:2017ccs,Huang:2017egk,Duan:2017qwj,Gu:2017bdw,Chao:2017yjg,Tang:2017lfb,Zu:2017dzm,Liu:2017rgs,Cao:2017ydw,Athron:2017drj,Gu:2017gle,Duan:2017pkq,Fang:2017tvj,Fan:2017sor,Yuan:2017ysv,Jin:2016kio,Okada:2017pgr,Sui:2017qra,Zhao:2017nrt,Ge:2017tkd}. In particular, dark matter (DM) was invoked to explain the excess. DM with a mass of about $1.4\tev$ could annihilate into electrons in a subhalo within about a kpc resulting in a narrow spike in the spectrum.

It is thus important to estimate the statistical significance of the excess. We do so with frequentist statistics in \refsec{sec:freq} and Bayesian statistics in \refsec{sec:bayes}. In each case, we fit the spectrum by three toy models:
\begin{itemize}
    \item A single power-law (PL),
    \begin{equation}
        \Phi(E) = \Phi_0 \left(\frac{E}{100\gev}\right)^{-p},
    \end{equation}
    described by a normalisation $\Phi_0$ and a power $p$.
    \item A smoothly-broken power-law (SBPL),
    \begin{equation}
       \begin{split}
           \Phi(E) = {}&{} \Phi_b \left(\frac{E}{100\gev}\right)^{-p_1} \times\\
                     {}&{} \left[1 + \left(\frac{E}{E_b}\right)^{(p_2 - p_1) / \Delta} \right]^{-\Delta},
       \end{split} 
    \end{equation}
    described by a normalisation $\Phi_b$, powers $p_1$ and $p_2$, a break $E_b$ and a smoothing parameter $\Delta$.
    This approximately equals two power-laws, which are smoothly matched at the break at $E_b$ by a smoothness governed by $\Delta$.
    \item A half-normal distribution upon a smoothly-broken power-law (signal),
    \begin{equation}
      \Phi(E) = \frac{A}{\sqrt{2\pi}\sigma}e^{-\frac{(E - m_\chi)^2}{2\sigma^2}}
    \end{equation}
    for $E \le m_\chi$ and zero elsewhere. This template is motivated by DM particles of mass $m_\chi$ annihilating into electrons in a nearby subhalo, resulting in a signal of amplitude $A$ and width $\sigma$.
\end{itemize}
The PL, SBPL and signal models have 2, 5 and 8 parameters, respectively. The toy models capture the behaviour of possible spectra from underlying physical processes. The unknown relationships between fundamental and toy model parameters cannot impact our frequentist analysis; however, they could influence suitable choices of prior in our Bayesian analysis. This is especially so for the width and amplitude of the signal, which could, in principle, be related to the DM annihilation cross section, subhalo properties and diffusion equations governing the propagation of charged cosmic rays.

DAMPE measured the average flux in 38 energy bins. We may predict the average flux in the $i$-th bin by
\begin{equation}
 \bar{\Phi}_i   \equiv \frac{1}{b_i - a_i} \int_{a_i}^{b_i} \Phi(E) \,\text{d}E,
\end{equation}
where the bin spans energies $a_i$ to $b_i$. DAMPE associated their measurement in the $i$-th bin with the energy $\langle E_i \rangle$ at which the predicted flux equals the predicted average flux in that bin for the best-fit SBPL model\cite{Lafferty:263644}, \ie $\langle E_i \rangle$ is defined by
\begin{equation}
    \Phi(\langle E_i \rangle) = \bar{\Phi}_i.
\end{equation}
The SBPL and PL fluxes are approximately linear on scales similar to the bin width such that $\Phi(\langle E_i \rangle) \approx \bar{\Phi}_i $ for the SBPL and PL models. The signal model, however, contains a peak that may be narrower than the bin width and we must explicitly calculate $\bar{\Phi}_i $ as it is not approximated by $\Phi(\langle E_i \rangle)$. This subtlety means that previous calculations of the required amplitude of a DM signal are underestimates by a factor of approximately the bin width divided by the signal width, $\Delta E/\sigma \approx \text{5 -- 20}$.

\section{Frequentist analysis}\label{sec:freq}

We performed two hypothesis tests: an SBPL versus a single PL under the hypothesis of a single PL, and an SBPL versus a signal under the hypothesis of an SBPL. We performed the former to validate our methodology against a result published by DAMPE. We used chi-squared test-statistics,
\begin{equation}
    \Delta \chi^2 = \min \chi^2(H_0) - \min \chi^2(H_1).
\end{equation}
We minimised the chi-squared with respect to each model's parameters with a CMA-ES evolutionary algorithm\cite{2016arXiv160400772H} implemented in \texttt{stochopy}\cite{stochopy}.  The chi-squared itself was
\begin{equation}
\chi^2 =\sum\limits_i \frac{\left(\bar\Phi_i - \mu_i\right)^2}{\sigma_i^2},
\end{equation}
where $\bar\Phi_i$ and $\mu_i$ were the predicted and measured average flux in the $i$-th bin, we summed over bins from $55\gev$ to $2.63\tev$ (matching the DAMPE analysis), and we added statistical and systematic errors in quadrature.

We found the distributions of our test-statistics by Monte Carlo. To do so, we generated 1000 pseudo-datasets from the best-fit single PL and best-fit SBPL models and reminimised the test-statistic for each dataset and model. Thus, we estimated the \pvalue,
\begin{equation}
    \pvalue = \cond{\Delta \chi^2 \ge \Delta \chi^2_\text{obs}}{H_0}
\end{equation}
by the fraction of pseudo-experiments in which the test-statistic exceeded that observed.  We, furthermore, calculated $68\%$ Clopper-Pearson intervals for the \pvalue \see{agresti2003categorical}.

\begin{figure}[tbp]
\centering 
\includegraphics[height=0.3\textwidth]{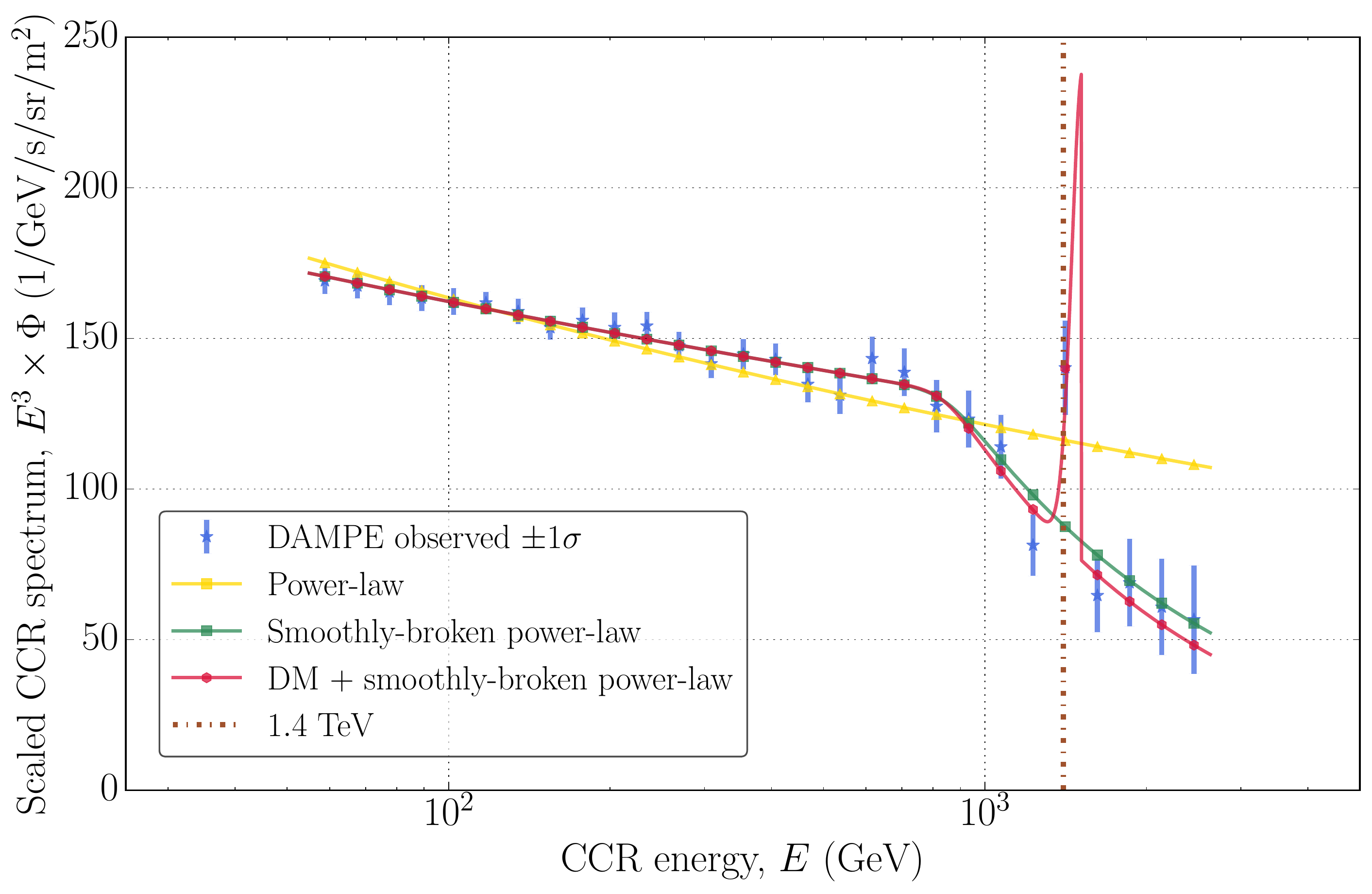}
\caption{Scaled energy spectrum of electrons and positrons measured by DAMPE (blue). Fits with a PL (yellow), SBPL (green) and an SBPL plus a DM signal (red) are also shown.}
\label{fig:spectrum}
\end{figure}

We found no differences in chi-squared between the PL and SBPL models as extreme as that observed in 1000 pseudo-experiments under the PL hypothesis. This resulted in a \pvalue associated with the PL model of at most $0.002$, which is equivalent to at least $2.9\sigma$. DAMPE applied Wilks' theorem to estimate the significance, finding $6.6\sigma$; however, in the limit $p_1 \to p_2$ the SBPL reduces to the single PL with no other parameters and, thus, Wilks' theorem cannot strictly apply. We found about $7\sigma$ with a similar procedure. Although we could not populate the tail of the distribution by Monte Carlo, since the observed test-statistic of about $56$ lies in the extreme tail of the distribution we expected that the \pvalue was negligible.  

Only 11 of our 1000 pseudo-experiments under the SBPL hypothesis had differences in chi-squared between the PL and SBPL models as extreme as that observed, resulting in a global significance of about $2.2\sigma$ -- $2.4\sigma$. This includes a two-dimensional look-elsewhere effect in the mass and width of the excess and corresponds to a \pvalue of about $1\%$. The local significance was about $3.6\sigma$, assuming a $\frac12\chi^2$ distribution for the test-statistic. To validate our methodology, we checked that our Monte Carlo reproduced a $\frac12\chi^2_1$ distribution from a model with a fixed mass and width. 

\begin{figure}[tbp]
\centering 
\includegraphics[height=0.3\textwidth]{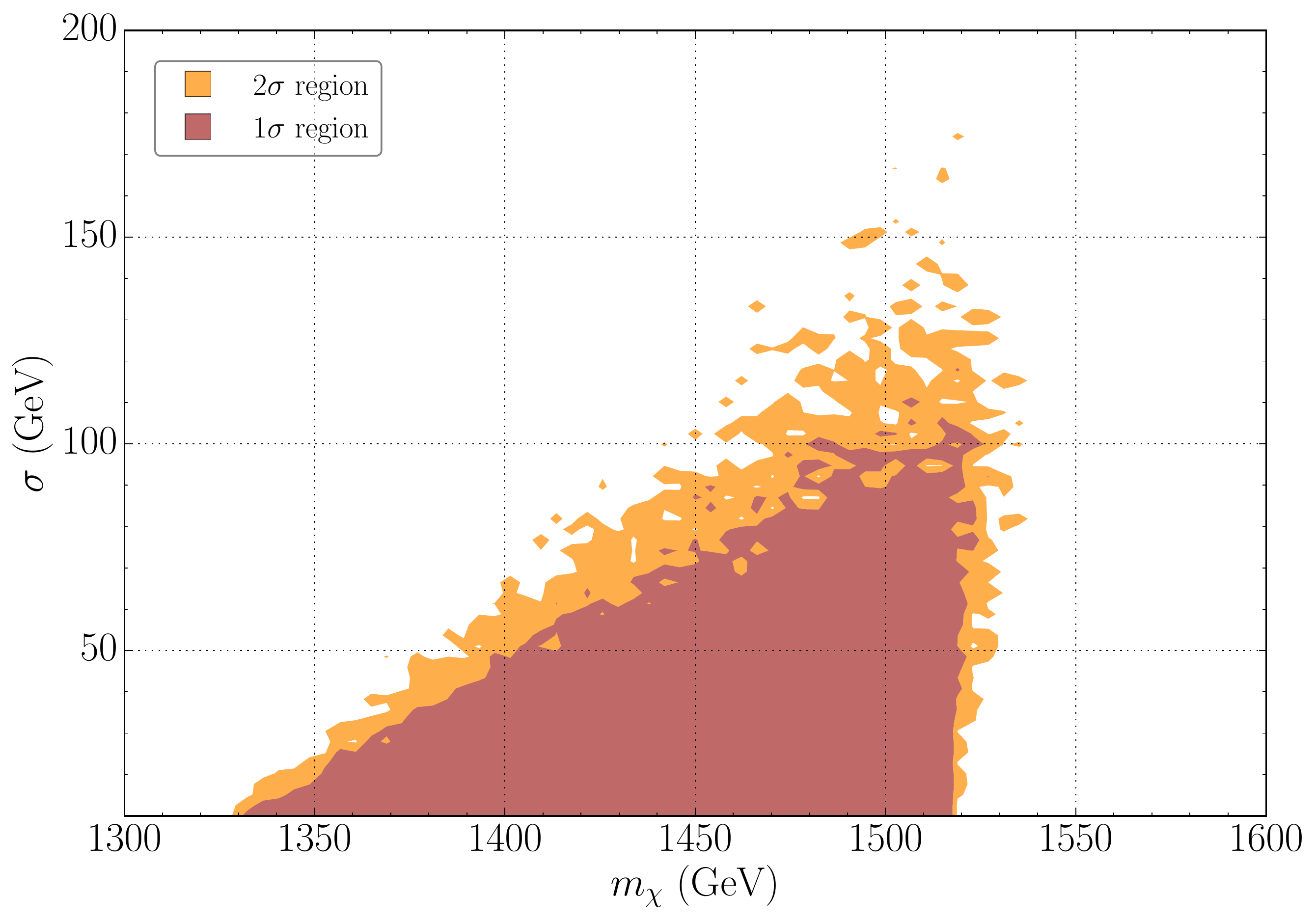}
\caption{Two-dimensional confidence interval for the DM mass and width of the DM signal.}
\label{fig:DM}
\end{figure}

We show best-fit spectra for our three models in \reffig{fig:spectrum}. There were degeneracies in the fits, especially in the amplitude and width of the signal. The amplitude of the narrow excess demonstrates that previous analyses underestimated the amplitude required to fit the anomalous bin. We show in \reffig{fig:DM}, furthermore, confidence regions for the DM mass and width of the signal. The DM signal must have a mass of about $1300\gev$ to $1500\gev$, a width of less than about $100\gev$, and an amplitude of about $10^{-5}\intflux$. This amplitude corresponds to a peak flux of about $10^{-7}\flux$ for a signal width of $\sigma = 10\gev$. 

\section{Bayesian analysis}\label{sec:bayes}

We considered Bayes factors between the three competing models of the spectrum. Bayes factors update the relative plausibility of two hypotheses with experimental data (see \refcite{Gregory});
\begin{equation}
    \text{Posterior odds} = \text{Bayes factor} \times \text{Prior odds}.
\end{equation}
The Bayes factor itself may be written
\begin{equation}
    B = \frac{\cond{D}{M_1}}{\cond{D}{M_2}},
\end{equation}
for data $D$, and models $M_1$ and $M_2$. This is a ratio of evidences,
\begin{equation}
    \cond{D}{M} = \int \cond{D}{M, x}\,\cond{x}{M} \,\text{d}x,
\end{equation}
where $x$ represents a model's parameters, $\cond{D}{M, x} = e^{-\frac12\chi^2}$ is our likelihood function and $\cond{x}{M}$ are our priors for the model's parameters. 
We calculated evidences with \texttt{(Py-)MultiNest-3.10}\cite{Buchner:2014nha,Feroz:2007kg,Feroz:2008xx,2013arXiv1306.2144F}. We list our priors in \reftable{tab:priors}. We picked flat priors for the exponents in the PL and SBPL models and logarithmic priors for all other parameters. Since we a priori knew the order of magnitude of the exponents, the choice of flat or logarithmic prior was moot. We found that, as anticipated, the SBPL model was favoured against the single PL model by about $10^{10}$. Since this was resounding and agreed with our frequentist analysis, we considered the matter settled and did not investigate prior sensitivity. 

\begin{table}
\begin{ruledtabular}
\begin{tabular}{lll}  
Parameter & Range & Prior\\
\hline
\multicolumn{3}{l}{Single power-law}\\
\hline
$\Phi_0$ & ($10^{-5}$ -- $10^{-3}$)\flux & Log\\
$p$ & 3 -- 4 & Linear\\
\hline
\multicolumn{3}{l}{Smoothly-broken power-law}\\
\hline
$\Phi_b$ & ($10^{-5}$ -- $10^{-3}$)\flux & Log\\
$p_1$ & 3 -- 4 & Linear\\
$p_2$ & 3 -- 5 & Linear\\
$E_b$ & (55 -- 2630)\gev & Log\\
$\Delta$ & $10^{-3}$ -- $1$ & Log\\
\hline
\multicolumn{3}{l}{DM Signal}\\
\hline
$A$ & ($10^{-7}$ -- $10^{-4}$)\intflux & Log\\
$m_\chi$ & (55 -- 2630)\gev & Log\\
$\sigma$ & (10 -- 500)\gev & Log\\
\end{tabular}
\end{ruledtabular}
\caption{\label{tab:priors}Priors for the model parameters in the Bayesian analysis of the DAMPE electron and positron spectrum.}
\end{table}

We found that the signal model was favoured versus an SBPL by a Bayes factor of about 2.  We anticipate that changes in priors for the SBPL parameters, which are present in each model, could not substantially modify the Bayes factor. We found that the Bayes factor increased to 4 with linear rather than logarithmic priors for the mass, amplitude and width of the DM signal. Our prior range for the amplitude spanned only three orders of magnitude about that favoured by the $1.4\tev$ excess and for the width spanned fewer than two orders of magnitude; arguably, they should have been more diffuse, which would decrease the Bayes factor. Our prior for the mass spanned the range searched by DAMPE, $55\gev$ to $2.63\tev$; shrinking it to between $1\tev$ to $2.63\tev$ could increase the Bayes factor to about 4. The maximum Bayes factor achievable with any priors is about $500$, which is obtained for Dirac delta functions at the best-fit mass, width and amplitude of a DM signal. Nevertheless, it seems difficult to make a reasonable case that the Bayes factor is compelling, especially since the narrow signal and substantial amplitude preferred by DAMPE were, if anything, a priori implausible as such a signal must originate from a nearby subhalo with a substantial DM density.
\vspace{0.3cm}

\section{Conclusions}\label{sec:concs}
The DAMPE energy spectrum of electrons and positrons contained two interesting features: a spectral break and a monochromatic excess. 
We performed a Bayesian and frequentist analysis of the features by testing three models: a single power-law, a smoothly-broken power-law, and a smoothly-broken power-law with a signal feature motivated by dark matter annihilation in a nearby subhalo. We found global \pvalue{}s through 1000 pseudo-experiments, including refits of models with 2, 5 and 8 parameters with evolutionary algorithms. We found Bayesian evidences by nested sampling. The break in the spectrum was significant with frequentist and Bayesian statistics --- we bounded the \pvalue at about $0.1\%$ and the Bayes factor was about $10^{10}$. We expect in fact that $\pvalue \lll 0.1\%$; our Monte Carlo may be unsuitable and specialised techniques such as Gross-Vitells\cite{Gross:2010qma} may be more appropriate. The excess, on the other hand, was present at $3.6\sigma$ local and $2.3\sigma$ global significance. The Bayes factor was sensitive to our choices of priors for the mass, amplitude and width of the signal, but for our choices favoured a signal by about $2$. Thus whilst intriguing, the excess is not currently compelling. We hope that this serves as a example of using frequentist and Bayesian methods for analysing anomalies in high-energy physics\cite{Fowlie:2016rmn}.

\bibliography{dampe}
\end{document}